\documentclass[aps,prl,a4paper,superscriptaddress,amsmath,amssymb,twocolumn]{revtex4} 
\usepackage{graphicx,epsfig}
\usepackage[usenames]{color}

\allowdisplaybreaks
\begin{document}

\newcommand{\Avg}[1]{\langle \,#1\,\rangle_G}

\newcommand{\E}{\mathcal{E}}
\newcommand{\Lag}{\mathcal{L}}
\newcommand{\M}{\mathcal{M}}
\newcommand{\N}{\mathcal{N}}
\newcommand{\U}{\mathcal{U}}
\newcommand{\R}{\mathcal{R}}
\newcommand{\F}{\mathcal{F}}
\newcommand{\V}{\mathcal{V}}
\newcommand{\C}{\mathcal{C}}
\newcommand{\I}{\mathcal{I}}
\newcommand{\s}{\sigma}
\newcommand{\up}{\uparrow}
\newcommand{\dw}{\downarrow}
\newcommand{\h}{\hat{\mathcal{H}}}
\newcommand{\Hn}{\mathcal{H}}
\newcommand{\himp}{\hat{h}}
\newcommand{\g}{\mathcal{G}^{-1}_0}
\newcommand{\D}{\mathcal{D}}
\newcommand{\A}{\mathcal{A}}
\newcommand{\projs}{\hat{\mathcal{S}}_d}
\newcommand{\proj}{\hat{\mathcal{P}}_d}
\newcommand{\K}{\textbf{k}}
\newcommand{\Q}{\textbf{q}}
\newcommand{\hzero}{\hat{T}}
\newcommand{\io}{i\omega_n}
\newcommand{\eps}{\varepsilon}
\newcommand{\+}{\dag}
\newcommand{\su}{\uparrow}
\newcommand{\giu}{\downarrow}
\newcommand{\0}[1]{\textbf{#1}}
\newcommand{\ca}{c^{\phantom{\dagger}}}
\newcommand{\cc}{c^\dagger}
\newcommand{\aaa}{a^{\phantom{\dagger}}}
\newcommand{\aac}{a^\dagger}
\newcommand{\bba}{b^{\phantom{\dagger}}}
\newcommand{\bbc}{b^\dagger}
\newcommand{\da}{d^{\phantom{\dagger}}}
\newcommand{\dc}{d^\dagger}
\newcommand{\fa}{f^{\phantom{\dagger}}}
\newcommand{\fc}{f^\dagger}
\newcommand{\ha}{h^{\phantom{\dagger}}}
\newcommand{\hc}{h^\dagger}
\newcommand{\be}{\begin{equation}}
\newcommand{\ee}{\end{equation}}
\newcommand{\bea}{\begin{eqnarray}}
\newcommand{\eea}{\end{eqnarray}}
\newcommand{\ba}{\begin{eqnarray*}}
\newcommand{\ea}{\end{eqnarray*}}
\newcommand{\dagga}{{\phantom{\dagger}}}
\newcommand{\bR}{\mathbf{R}}
\newcommand{\bQ}{\mathbf{Q}}
\newcommand{\bq}{\mathbf{q}}
\newcommand{\bqp}{\mathbf{q'}}
\newcommand{\bk}{\mathbf{k}}
\newcommand{\bh}{\mathbf{h}}
\newcommand{\bkp}{\mathbf{k'}}
\newcommand{\bp}{\mathbf{p}}
\newcommand{\bL}{\mathbf{L}}
\newcommand{\bRp}{\mathbf{R'}}
\newcommand{\bx}{\mathbf{x}}
\newcommand{\by}{\mathbf{y}}
\newcommand{\bz}{\mathbf{z}}
\newcommand{\br}{\mathbf{r}}
\newcommand{\Ima}{{\Im m}}
\newcommand{\Rea}{{\Re e}}
\newcommand{\Pj}[2]{|#1\rangle\langle #2|}
\newcommand{\ket}[1]{\vert#1\rangle}
\newcommand{\bra}[1]{\langle#1\vert}
\newcommand{\setof}[1]{\left\{#1\right\}}
\newcommand{\fract}[2]{\frac{\displaystyle #1}{\displaystyle #2}}
\newcommand{\Av}[2]{\langle #1|\,#2\,|#1\rangle}
\newcommand{\av}[1]{\langle #1 \rangle}
\newcommand{\Mel}[3]{\langle #1|#2\,|#3\rangle}
\newcommand{\Avs}[1]{\langle \,#1\,\rangle_0}
\newcommand{\eqn}[1]{(\ref{#1})}
\newcommand{\Tr}{\mathrm{Tr}}

\newcommand{\Pg}{\mathcal{P}_G}

\newcommand{\Vb}{\bar{\mathcal{V}}}
\newcommand{\Vd}{\Delta\mathcal{V}}
\newcommand{\Pb}{\bar{P}_{02}}
\newcommand{\Pd}{\Delta P_{02}}
\newcommand{\tb}{\bar{\theta}_{02}}
\newcommand{\td}{\Delta \theta_{02}}
\newcommand{\Rb}{\bar{R}}
\newcommand{\Rd}{\Delta R}

\title{Energetics and Electronic Structure of Plutonium}

\author{Nicola Lanat\`a}
\altaffiliation{Equally contributed to this work}
\affiliation{Department of Physics and Astronomy, Rutgers University, Piscataway, New Jersey 08856-8019, USA}
\author{Yong-Xin Yao}
\altaffiliation{Equally contributed to this work}
\affiliation{Ames Laboratory-U.S. DOE and Department of Physics and Astronomy, Iowa State
University, Ames, Iowa IA 50011, USA}
\author{Cai-Zhuang Wang}
\affiliation{Ames Laboratory-U.S. DOE and Department of Physics and Astronomy, Iowa State
University, Ames, Iowa IA 50011, USA}
\author{Kai-Ming~Ho}
\affiliation{Ames Laboratory-U.S. DOE and Department of Physics and Astronomy, Iowa State
University, Ames, Iowa IA 50011, USA}
\author{Gabriel Kotliar}
\affiliation{Department of Physics and Astronomy, Rutgers University, Piscataway, New Jersey 08856-8019, USA}

\date{\today} 

\maketitle

{\bf 
Plutonium is the most exotic and mysterious element in the
periodic table. It has $6$ metallic phases and
peculiar physical properties not yet understood.
One of the most intriguing properties of Pu
is that relatively small changes of temperature
can induce transitions between different structures,
that are accompanied by very large changes of equilibrium volumes.
This fact
has stimulated extensive
theoretical and experimental studies.
In spite of this, a convincing explanation
of the metallurgic properties of Pu
based on fundamental principles is still lacking,
and none of the previous theories has been
able to describe simultaneously the energetics and the $f$ electronic
structure of all of the phases of Pu on the same footing.
Here we provide a bird's eye view of Pu by
studying the zero-temperature pressure-volume phase diagram
and the $f$ electronic structure 
of all of its crystalline phases 
from first principles.
In particular, we clarify the way in which
the $f$-electron correlations determine its unusual energetics.
Our theoretical energetics and ground-state $f$ electronic structure
are both in good quantitative agreement with the experiments. 
}

\begin{figure}
  \begin{center}
    \includegraphics[width=8.2cm]{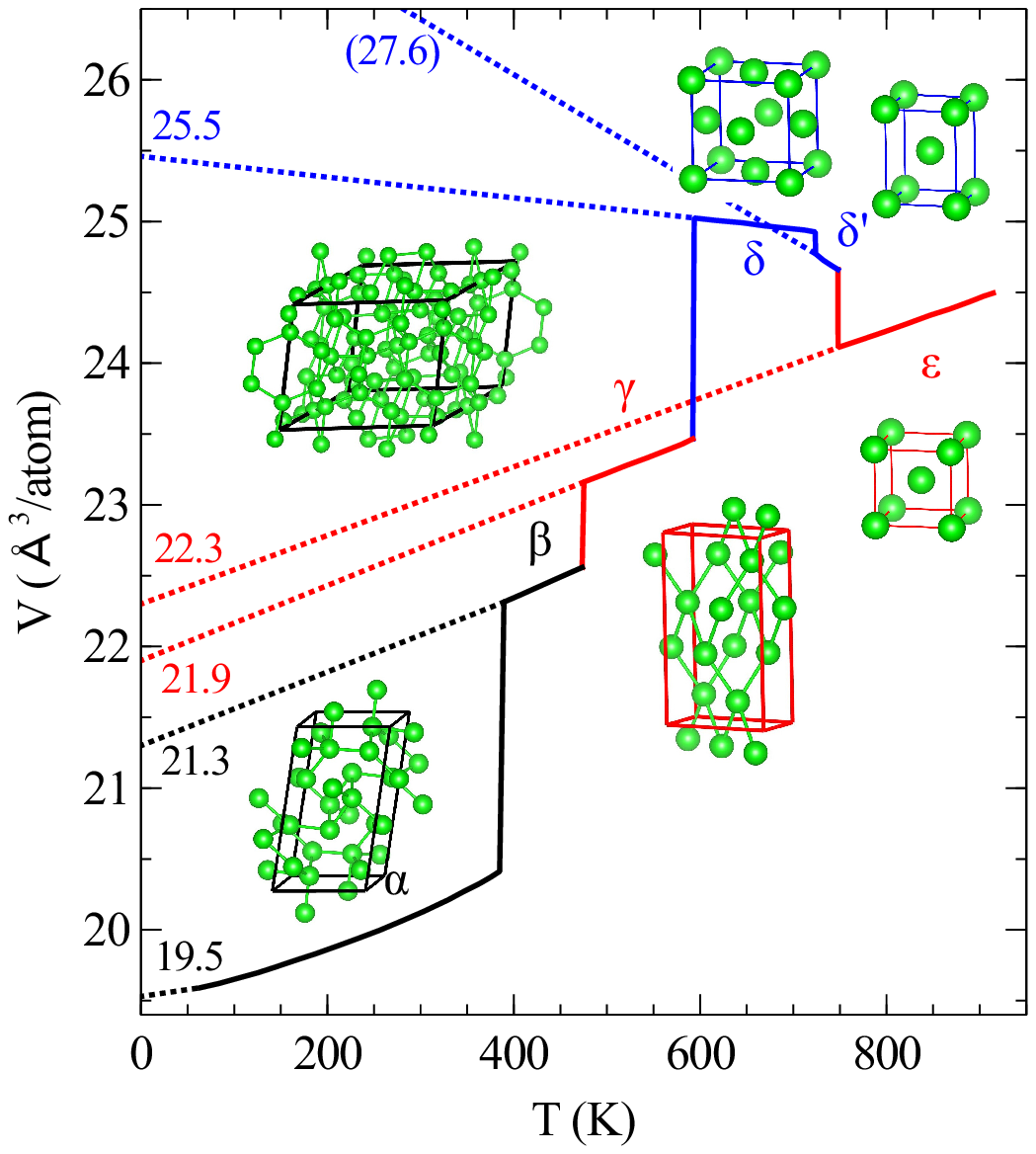}
    \caption{Experimental volume-temperature phase-diagram of Pu.
    The dotted lines indicate the zero-temperature equilibrium
    volumes extrapolated by linear interpolation.}
    \label{figure1}
  \end{center}
\end{figure}

The stable structure of plutonium at ambient conditions is 
$\alpha$-Pu, that has a low-symmetry monoclinic structure with 
$16$ atoms within the unit cell 
grouped in $8$ nonequivalent types.
At higher temperatures, see Fig.~\ref{figure1},
Pu can assume the following distinct lattice structures:
$\beta$ (monoclinic, with $34$ atoms within the unit cell grouped in $7$
inequivalent types), 
$\gamma$ (orthorhombic), 
$\delta$ (fcc), $\delta'$ (bct), and $\epsilon$ (bcc).
These temperature-induced structure-transitions are accompanied
by significant changes of volume, that are still not well understood.  
In particular,
the equilibrium volumes of the $\delta$ and $\delta'$ phases
are very large with respect to the other allotropes.
Another interesting puzzle is that $\delta$- and $\delta'$-Pu 
have negative thermal-expansion coefficients within their respective
range of stability, 
unlike the vast majority of materials.

It is clear that a theoretical explanation of the
properties of Pu
needs to be supported by first-principles
calculations which, not only are able to take into account
the details of the band-structure and the strong-correlation effects,
but are also able to evaluate precisely the total energy.
Previous state-of-the-art density functional theory
(DFT) calculations~\cite{dft-like-2,dft-like,dft-like-3,dft-like-4} were able to reproduce the energetics of Pu, but
in order to describe all of the phases on the same footing
it was necessary to introduce artificial~\cite{Absence_magnetic_moments_Pu}
spin and/or orbital polarizations
--- thus compromising the description of the electronic structure.
Calculations within the framework of
DFT in combination with dynamical
mean field theory (DFT+DMFT) have been able to explain
several aspects of the electronic structure of Pu,
see, e.g., Refs.~\cite{Pu-Gabi-nature-2001,Pu-nature-mixed_valence,sh-alpha-delta-Pu,deltaPu-ImportanceFullCoulomb-1}.
Nevertheless, the computational complexity of this approach made it impossible
to calculate the pressure-volume
phase diagram of all of the phases of Pu.
In this work we overcome these issues
by using the charge self-consistent combination of
the Local Density Approximation and the Gutzwiller Approximation
(LDA+GA)~\cite{Gutzwiller3,Zein,Fang,Ho,Gmethod},
whose description of the ground-state properties is generally
in very good agreement with LDA+DMFT,
but is considerably less computationally demanding.
In particular, we employ the numerical approach derived
in Ref.~\cite{DMFT-GA}.

\begin{figure*}[ht]
\begin{center}
\includegraphics[width=16.0cm]{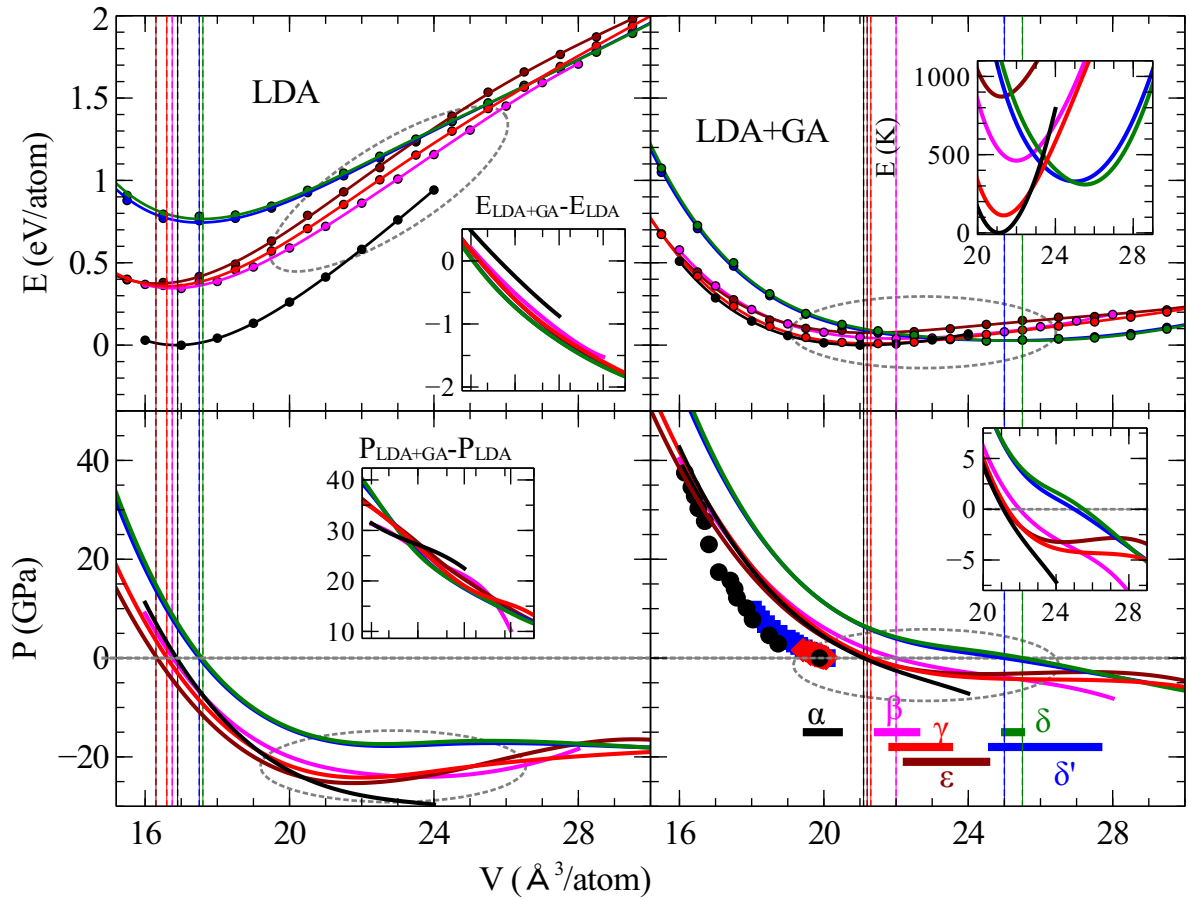}
\caption{Theoretical total energies for the crystalline phases of Pu
  as a function of the volume (upper panels)
  and corresponding pressure-volume curves (lower panels).
  Our results are shown both in LDA (left panels) and in LDA+GA
  (right panels).
  The right insets are zooms of the curves in the corresponding panels.
  In the upper-left inset are shown the correlation energies
  --- here defined as the differences between the LDA+GA and LDA
  total energies, ---
  while the corresponding contributions
  to the pressure are shown in the lower-left inset.
  The vertical lines indicate the minima of the energy curves.
  The lines of the legend indicate the estimated
  zero-temperature equilibrium volumes.
  The LDA+GA pressure-volume curves are shown
  in comparison with the experimental data of $\alpha$-Pu from
  Ref.~\cite{ExpPublack}
  (black circles), Ref.~\cite{ExpPublue} (blue squares)
  and Ref.~\cite{ExpPured} (red diamonds).
}
\label{figure2}
\end{center}
\end{figure*}

\begin{table}
  \begin{center}      
    \caption{Zero-temperature theoretical equilibrium volumes, bulk modulus
      and total energies of the crystalline phases of Pu
      in comparison with the experiments~\cite{elastic-Pu,energies_Pu_allotropes}.
      The total energies are relative to the ground-state energy of $\alpha$-Pu.
      The zero-temperature equilibrium volumes of $\gamma$- and $\epsilon$-Pu
      are extrapolated by linear interpolation from the experimental data
      reported in Fig.~\ref{figure1}.
      Note that the experimental value of the $\delta$-Pu bulk-modulus
      refers to the $1.9\%$ Ga alloy.}
    \begin{ruledtabular}
      \begin{tabular}{lllllll}
        Pu  & $\alpha$ & $\beta$ & $\gamma$ & $\delta$ & $\delta'$ & $\epsilon$ \\
        \\\hline
        $V_{\text{exp}}\,(\AA^3)$    & 19.5 & 21.9 & 22.3 & 25.2 & 25.1 & 22.3  \\
        $V_{\text{th}}\pm 0.5\,(\AA^3)$ & 21.1 & 22.0  & 21.3 & 25.5 & 25.0 & 21.2  \\
        \hline
        $E_{\text{exp}}\,(K)$ & 0 & 470 & 550 & 620 & 710 & 1130  \\
        $E_{\text{th}}\pm 100\,(K)$ & 0 & 460 & 110 & 310 & 330 & 870  \\
        \hline
        $K_{\text{exp}}\,(GPa)$   & 70.2 & --- & --- & 38 & --- & ---  \\
        $K_{\text{th}}\,(GPa)$  & 50--70 & 40--55 & 45--70 & 15--35 & 10--25 & 35--55
      \end{tabular}
      \label{table1}
    \end{ruledtabular}
  \end{center}
\end{table}

In the upper panels of Fig.~\ref{figure2} are shown the LDA (left) and
LDA+GA (right) evolutions of the total energies $\mathcal{E}$ as a function
of the volume $V$ for all of the crystalline phases of Pu. 
In the lower panels are reported the corresponding evolutions of the
pressure $\mathcal{P}=-d\mathcal{E}/dV$ in comparison with the experimental data of $\alpha$-Pu.
In the left insets are shown the correlation energies 
--- here defined as the differences between the LDA+GA and LDA 
total energies ---
and the respective contributions to the pressure.
Note that in our calculations we have not performed structure relaxation,
but we have assumed a uniform rescaling of the experimental
lattice parameters, see Refs.~\cite{ExpPured,beta-Pu-struct-exp}.

In table~\ref{table1} the theoretical zero-temperature
equilibrium volumes are shown in comparison with the zero-temperature
experimental volumes, that we assume to be in between the
thermal-equilibrium volumes and the zero-temperature
values extrapolated by linear interpolation in Fig.~\ref{figure1}.
The bulk modulus and energies
(referred to the ground-state energy of $\alpha$-Pu)
are shown in comparison with the experimental data
of Refs.~\cite{elastic-Pu,energies_Pu_allotropes}.
Note that the numerical error for the theoretical bulk modulus is due to the
fact our calculations are performed only on a discrete mesh of values,
see the upper panels of Fig.~\ref{figure2}.
Remarkably,
while the theoretical equilibrium volumes of all phases of Pu are
very similar in LDA, they are very different in LDA+GA, and
in good quantitative agreement with the zero-temperature
experimental values.
Furthermore, while LDA predicts very large equilibrium 
energy-differences between the phases of Pu,
these differences are very small in LDA+GA, in agreement
with the experiments. Note also that the LDA+GA ground-state
energies increase monotonically from each phase to the
next-higher-temperature phase,
consistently with the experiments, see 
Fig.~\ref{figure1} and Table~\ref{table1}.
The only exception is $\beta$-Pu, whose theoretical equilibrium energy
is larger than $\gamma$-, $\delta$- and $\delta'$-Pu.

In order to understand how the electron-correlations affect
so drastically the energetics of Pu,
it is enlightening to look at
the behavior of the correlation energies, see the left insets in Fig.~\ref{figure2}.
In fact, the evolution of the correlation energies as a function of the volume
is essentially structureless and identical for all of the
phases (made exception for a uniform structure-dependent
energy shift whose main effect
is to slightly increase the energy of $\alpha$-Pu with respect to the
other phases).
As a result of this correction,
the LDA total energies 
are transformed
in the way indicated by the gray circles
in the upper panels of Fig.~\ref{figure2}.
The relative behavior of the LDA+GA zero-temperature energies
of the allotropes of Pu is clearly 
inherited by the LDA energy-volume curves in the
region highlighted in the upper-left panel of Fig.~\ref{figure2},
which transform into the region highlighted in the upper-right panel
of Fig.~\ref{figure2} when the 
correlation energies are taken into account.
The same considerations apply to the evolutions of the pressure,
as indicated by the gray circles
in the lower panels of Fig.~\ref{figure2}.
The above observation explains from a simple perspective 
how the electron
correlations determine the unusual energetics of Pu.

\begin{figure}
  \begin{center}
    \includegraphics[width=8.7cm]{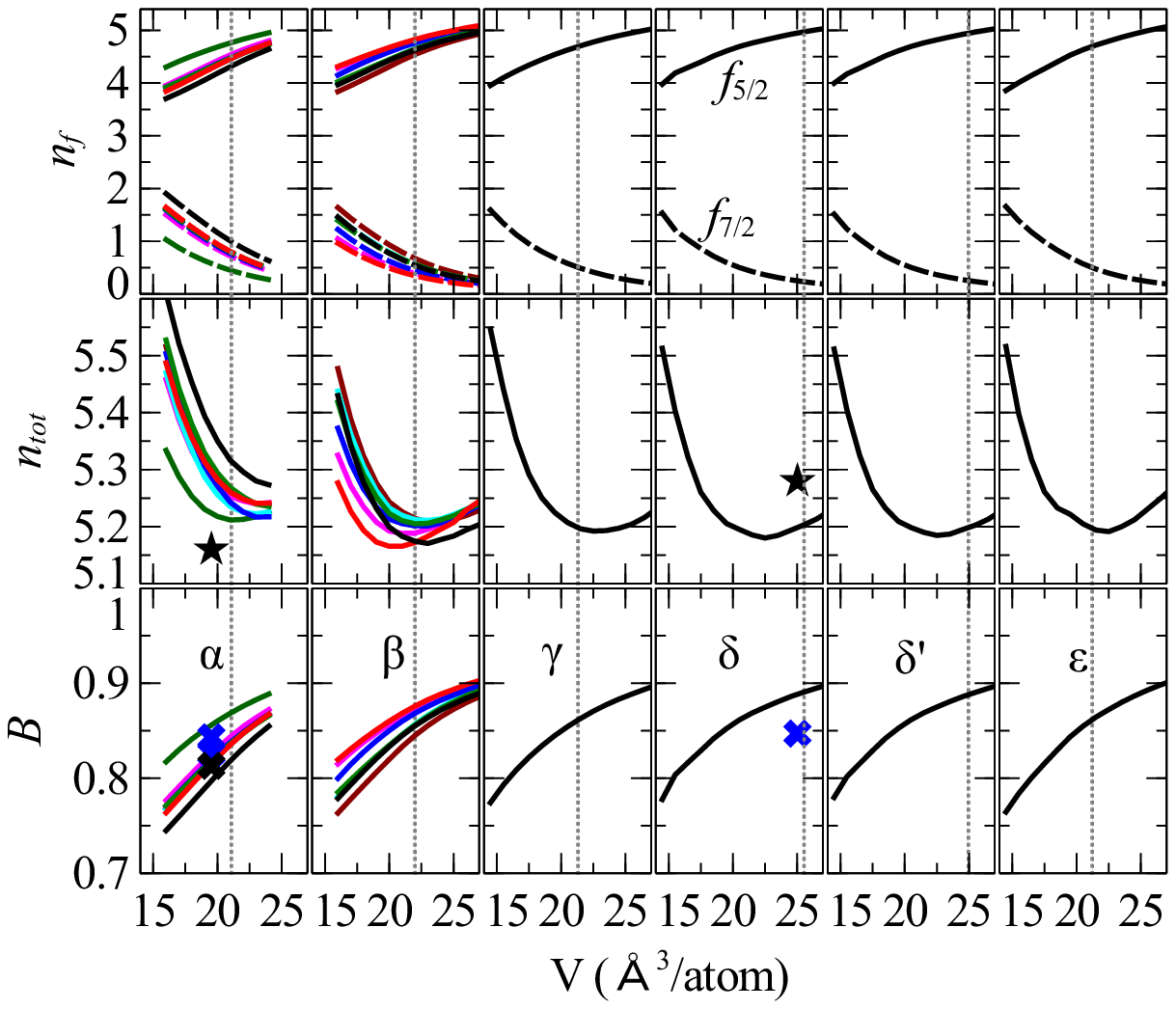}
    \caption{Upper panels: evolution as a function of the volume
      of the averaged orbital populations 
      of the $5/2$ and $7/2$ $f$ electrons.
      Middle panels: total orbital occupations
      in comparison
      with the values extrapolated in Ref.~\cite{pnas-Pu-fn} 
      from XANES measurements at ambient conditions 
      of $\alpha$-Pu (black stars) and the $1.9\%$-Ga $\delta$-Pu alloy.
      Lower panels: theoretical branching ratios
      in comparison with the values extrapolated in
      Refs.~\cite{Branching-ratio_EXP,Branching-ratio_PRB}
      from XAS (black cross) and EELS (blue crosses)
      experiments of $\alpha$-Pu and the $0.6\%$-Ga $\delta$-Pu alloy.
      The colors in the first and second panels from the left
      correspond to the 
      inequivalent atoms of $\alpha$-Pu and $\beta$-Pu.
      The vertical dotted lines indicates the LDA+GA equilibrium
      volumes for the respective phases. 
    }
    \label{figure3}
  \end{center}
\end{figure}
In the upper panels of Fig.~\ref{figure3} are shown
the occupations of the $f$ electrons.
The total number of $f$ electrons in $\delta$-Pu is $n_f\simeq 5.2$,
which is consistent with previous LDA+DMFT calculations~\cite{Pu-nature-mixed_valence}.
Here we find that $n_f\simeq 5.2$
also for $\gamma$-, $\delta'$- and $\epsilon$-Pu.
In the monoclinic structures $n_f$ is different for the inequivalent atoms
within the unit cell, and
it runs between $5.21$ and $5.32$ in $\alpha$-Pu
while
it runs between $5.17$ and $5.21$ in $\beta$-Pu.
In the middle panels of Fig.~\ref{figure3} are shown the averaged orbital
populations with total angular momentum $J=7/2$ and $J=5/2$.
Note that for all of the phases of Pu the number of $7/2$ $f$ electrons decreases
as a function of the volume, while the number of $5/2$ $f$ electrons
increases.
This behavior indicates simply that the spin-orbit effect
is more effective at larger volumes, as expected.
Finally, in the lower panels is shown the behavior of the
branching ratio $B$, which is a measure of the strength
of the spin-orbit coupling interaction in the
$f$ shell, and is calculated from the orbital populations
making use of the equation
\be
B=\frac{3}{5}-\frac{4}{15}\frac{1}{14-n_{5/2}-n_{7/2}}
\left(\frac{3}{2}\,n_{7/2}-2\,n_{5/2}\right)\,,
\ee
see Ref.~\cite{Branching-ratio_PRX,Branching-ration_EPL}.
Consistently with the behavior of the orbital populations,
$B$ increases as a function of the volume.
Note that the behavior of $n_f$ and $B$ is very similar
for all of the phases of Pu.

The $\alpha$-Pu theoretical value of $n_f$ at equilibrium is 
in good agreement with the values extrapolated from the 
$X$-ray absorption near-edge structure (XANES)
measurements of Ref.~\onlinecite{pnas-Pu-fn}.
On the other hand, while our calculations indicate that 
$n_f$ is slightly smaller in $\delta$-Pu than in $\alpha$-Pu,
according to the extrapolations of Ref.~\onlinecite{pnas-Pu-fn},
the $1.9\%$-Ga $\delta$-Pu alloy has a larger $n_f$ with respect
to $\alpha$-Pu.
Also the theoretical values of $B$
are in good agreement with values extrapolated
in Refs.~\cite{Branching-ratio_EXP,Branching-ratio_PRB}
from electron energy-loss spectroscopy (EELS) and
X-ray absorption spectroscopy (XAS)~\cite{XAS1,XAS2,XAS3,XAS4}.

Let us study the behavior of the many-body reduced density matrix
$\hat{\rho}_f$ of the $f$ electrons, which is obtained from the full
many-body density matrix of the system by tracing out all of the degrees
of freedom with the exception of the $f$ local many-body configurations
of one of the Pu atoms.
We define
\be
\hat{F}\equiv -\ln\hat{\rho}_f + k\,,
\ee
where $k$ is an arbitrary constant that we determine
so that the lowest eigenvalue of $\hat{F}$ is zero by definition.
Within this definition $\hat{\rho}_f\propto e^{-\hat{F}}$, i.e., $\hat{F}$
represents an effective local Hamiltonian of the $f$ electrons
that depends on the volume, and that is renormalized with respect
to the atomic $f$ Hamiltonian because of the entanglement
with the rest of the system, see Ref.~\cite{pmee}.

\begin{figure*}[ht]
\begin{center}
\includegraphics[width=18.0cm]{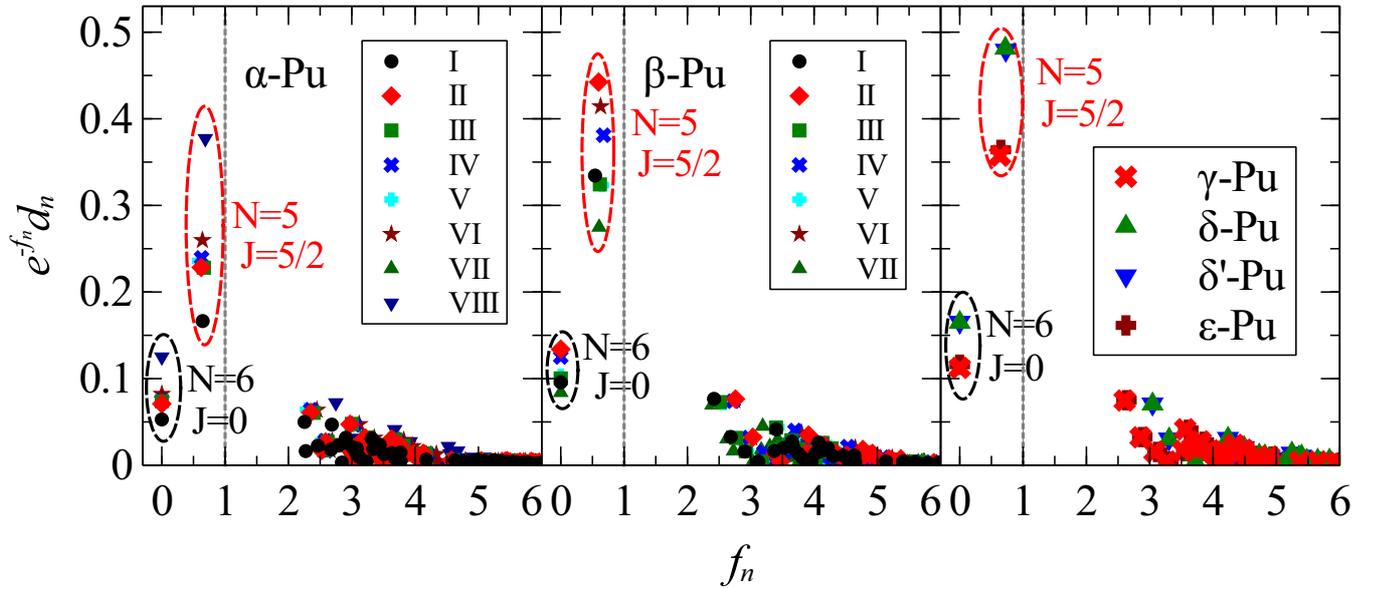}
\caption{Configuration probabilities of the eigenstates of the
reduced density matrix $\hat{\rho}_f\equiv e^{-\hat{F}}/\Tr[e^{-\hat{F}}]$
of the $f$ electrons as a function of
the eigenvalues $f_n$ of $\hat{F}$.
Each configuration probability is weighted by the degeneracy
$d_n=2J_n+1$ of the respective eigenvalue $f_n$,
where $J_n$ is the total angular momentum.
}
\label{figure4}
\end{center}
\end{figure*}
Fig.~\ref{figure4} shows the eigenvalues $P_n$
of $\hat{\rho}_f$ as a function of the eigenvalues $f_n$ of $\hat{F}$
for all of the allotropes of Pu
(that are computed at their respective theoretical zero-temperature
equilibrium volumes).
Consistently with Ref.~\cite{Pu-nature-mixed_valence},
we find that for $\delta$-Pu
there are two dominant groups of multiplets: one with $N=5$ and $J=5/2$
(that is $6$ times degenerate), 
and one with $N=6$ and $J=0$
(that is non-degenerate).
Interestingly, our results show that this conclusion
applies also to all of the other phases.
The $f$ probability distribution of $\delta$- and $\delta'$-Pu
is slightly less broad with respect to the other phases.
This is to be expected, as $\delta$- and $\delta'$-Pu are stable
at larger volumes, so that the local $f$ degrees of freedom
are less entangled with the rest of the system with respect to the
other phases.
The $f$ probability distributions of $\alpha$- and $\beta$-Pu are considerably
different for inequivalent atoms, as they depend on the number
and relative distances of the nearest-neighbor atomic positions,
consistently with Refs.~\cite{alphaPu-Gabi-nature-2013,DMFT-GA}.
Note that in $\beta$-Pu the atom-dependency of the $f$
probability distribution is less pronounced than in $\alpha$-Pu.

We point out that the fact that $n_f\simeq 5.2$ reveals
that the $f$ electrons of Pu are in a pronounced
mixed-valence state~\cite{Mixed-valence_Varma}.
Indeed, the probability of the $N=6,J=0$ multiplet is very large,
as indicated by the fact that it has
the lowest $\hat{F}$ eigenvalue.
The reason why $n_f$ is closer to $5$ than to $6$ is that
the $N=6,J=0$ multiplet is non-degenerate, while the 
$N=5,J=5/2$ $\hat{F}$-eigenvalue is $6$ times degenerate
--- so that its contribution to $n_f$ is weighted by a
factor $6=2\times 5/2 + 1$.
The observation that the $f$ electrons have 
a significant mixed-valence character 
indicates that the local $f$ degrees of freedom are highly entangled
with the rest of the system~\cite{pmee}.
This observation is consistent with the
fact that the Pauli susceptibility of the $\delta$-Pu Ga alloy is Pauli-like
at low temperatures~\cite{Absence_magnetic_moments_Pu}.
Furthermore, it is consistent with the
statement of Ref.~\cite{energies_Pu_allotropes} that Pu is an ordinary
quasiharmonic crystal in all of its crystalline phases, i.e.,
that already at $T\gtrsim 200\,K$ the electronic entropy is very small with
respect to the quasiharmonic contributions.

In conclusion, in this work we have calculated from first principles
the zero-temperature energetics of Pu,
finding very good agreement with the experiments.
Our analysis has clarified how the electron
correlations determine the unusual energetics of Pu, including
the fact that the different allotropes have very large equilibrium-volume
differences while they are very close in energy.
Remarkably, in our calculations we did not
introduce any artificial
spin and/or orbital polarizations~\cite{Absence_magnetic_moments_Pu},
while this was necessary in previous state-of-the-art
DFT calculations~\cite{dft-like-2,dft-like,dft-like-3,dft-like-4}.
This advancement has enabled us to 
describe also the $f$ electronic structure of Pu
on the same footing.
Our calculations indicate that the ground-state 
$f$ electronic structure is similar for all phases of Pu, and
that the $f$-electron atomic probabilities
display a significant mixed-valence character.
Our zero-temperature calculations of Pu constitute also
an important step toward
the theoretical understanding of its peculiar
temperature-dependent properties, e.g.,
the negative thermal expansion of $\delta$- and $\delta'$-Pu.
In fact, above room temperature, 
the contributions to the free energy of the
non-adiabatic effects and of the thermal excitations
of the electrons from their ground state are expected to be negligible
in Pu~\cite{energies_Pu_allotropes}.
Consequently, the ground-state total energy could be used to 
derive microscopically the atomistic potentials and
study from first principles
the evolution of the atom positions as a function of
the temperature making use of the Born-Oppenheimer approximation.

\section{Methods}

In this work we have employed the LDA+GA 
charge self-consistent numerical scheme of
Ref.~\cite{DMFT-GA}. The interface between Wien2k~\cite{w2k} and
the GA has been coded following Ref.~\cite{Haule10}.
The calculations have been performed making use of
the general Slater-Condon parametrization of the on-site interaction,
assuming $U=4.5\,eV$ and $J=0.36\,eV$, consistently with Ref.~\cite{DMFT-GA}.
We have employed the following 
standard form for the double-counting functional
\be
E_{\text{dc}}^{\text{St}}[n_f]=
\frac{U}{2}\,n_f(n_f-1)-\frac{J}{2}\,n_{f}(n_{f}/2-1)
\,,
\ee
where $n_f\equiv\Av{\Psi}{\hat{n}_f}$ is the total number
of $f$ electrons.

\section{Contributions}

~N.L. and Y.X.Y.
carried out the LDA+GA calculations, analyzed the data, and
co-developed the GA code (open source ``FastGutz'' package)
under the supervision of G.K., C.Z.W and K.M.H. .
~N.L. led the project and wrote the manuscript.
~Y.X.Y. coded the interface between Wien2k and the GA solver, which was
constructed on the basis of the LDA+DMFT interface developed by K.H., and
parallelized the GA solver. G.K. proposed the project and supervised
the research.

\section{Acknowledgments}

We thank XiaoYu Deng and Kristjan Haule for useful discussions.
N.L. and G.K. were supported by U.S. DOE Office of
Basic Energy Sciences under Grant No. DE-FG02-99ER45761.
The collaboration was supported by the U.S. Department of Energy
through the Computational Materials and Chemical Sciences Network CMSCN.
Research at Ames Laboratory supported by the U.S.
Department of Energy, Office of Basic Energy Sciences,
Division of Materials Sciences and Engineering. 
Ames Laboratory is operated for the U.S. Department of Energy
by Iowa State University under Contract No. DE-AC02-07CH11358.

\bibliography{plutonium}

\end{document}